\documentclass[conference]{IEEEtran}

\usepackage{microtype}

\usepackage[utf8]{inputenc}

\usepackage[T1]{fontenc}
\usepackage{amsmath}
\usepackage{bm}
\usepackage{xspace}

\usepackage[english]{babel}

\usepackage{hyperref}

\usepackage{todonotes}

\usepackage{siunitx}

\usepackage{subfig}  

\usepackage{dblfloatfix}

\usepackage{blindtext}

\usepackage{listings}
\usepackage{booktabs}
\usepackage{color, colortbl}
\definecolor{Gray}{gray}{0.9}
\definecolor{LightGray}{gray}{0.97}
\usepackage{marvosym}

\usepackage{fancyhdr}
\title{Q-Learning Inspired Self-Tuning\\for Energy Efficiency in HPC}

\author{
    \IEEEauthorblockN{Andreas Gocht \href{https://orcid.org/0000-0003-2760-3672}{\includegraphics[height=9pt,keepaspectratio]{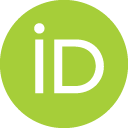}}, Robert Sch\"{o}ne, Mario Bielert}
    \IEEEauthorblockA{
        \Letter \, Center for Information Services and High Performance Computing (ZIH)\\
        Technische Universit\"{a}t Dresden, 01062 Dresden, Germany \\
        \MVAt \, \{andreas.gocht | robert.schoene | mario.bielert\}@tu-dresden.de 
    }
}

\begin{document}


\ifx\DraftModeOn\undefined

\newcommand{\todoti}[1]{}
\newcommand{\todomb}[1]{}
\newcommand{\todors}[1]{}
\newcommand{\todoag}[1]{}
\newcommand{\tododh}[1]{}

\newcommand{\figref}[1]{Figure~\ref{fig:#1}}
\newcommand{\tabref}[1]{Table~\ref{tab:#1}}
\newcommand{\secref}[1]{Section~\ref{sec:#1}}
\newcommand{\lstref}[1]{Algorithm~\ref{alg:#1}}

\newcommand{\papertarget}[2]{}

\renewcommand{\todo}[1]{}

\else

\newcommand{\todoti}[1]{\todo[color=yellow!60,inline,size=\small]{Thomas: #1}}
\newcommand{\todomb}[1]{\todo[color=cyan!60,inline,size=\small]{Mario: #1}}
\newcommand{\todors}[1]{\todo[color=green!60,inline,size=\small]{Robert: #1}}
\newcommand{\todoag}[1]{\todo[color=orange!60,inline,size=\small]{Andreas: #1}}
\newcommand{\tododh}[1]{\todo[color=red!60,inline,size=\small]{Daniel: #1}}

\newcommand{\figref}[1]{\textcolor{red}{Figure~\ref{fig:#1}}}
\newcommand{\tabref}[1]{\textcolor{red}{Table~\ref{tab:#1}}}
\newcommand{\secref}[1]{\textcolor{red}{Section~\ref{sec:#1}}}
\newcommand{\lstref}[1]{\textcolor{red}{Algorithm~\ref{alg:#1}}}

\newcommand{\papertarget}[2]{\todo[inline]{Target: #1: \\Deadline: #2}}

\fi

\maketitle


\cfoot{\footnotesize \textbf{© 2019 IEEE}. Personal use of this material is permitted. Permission from IEEE must be obtained for all other uses, in any current or future media, including reprinting/republishing this material for advertising or promotional purposes, creating new collective works, for resale or redistribution to servers or lists, or reuse of any copyrighted component of this work in other works. The definite version is available at \url{https://doi.org/10.1109/HPCS48598.2019.9188112}.}
\renewcommand{\headrulewidth}{0pt}
\thispagestyle{fancy}

\begin{abstract}
System self-tuning is a crucial task to lower the energy consumption of computers.
Traditional approaches decrease the processor frequency in idle or synchronisation periods.
However, in High-Performance Computing (HPC) this is not sufficient: if the executed code is load balanced, there are neither idle nor synchronisation phases that can be exploited.
Therefore, alternative self-tuning approaches are needed, which allow exploiting  different compute characteristics of HPC programs.

The novel notion of application regions based on function call stacks, introduced in the Horizon 2020 Project READEX, allows us to define such a self-tuning approach.
In this paper, we combine these regions with the Q-Learning typical state-action maps, which save information about available states, possible actions to take, and the expected rewards.
By exploiting the existing processor power interface, we are able to provide direct feedback to the learning process.
This approach allows us to save up to 15\% energy, while only adding a minor runtime overhead.
\end{abstract}

\vspace{-0.4em}

\vspace{-0,4em}
\section{Introduction}
\label{sec:intro}
\vspace{-0.4em}

%

Optimising parallel software for energy by using techniques like Dynamic Voltage and Frequency Scaling (DVFS) is a broadly studied topic.
Researchers introduced different concepts to lower the power consumption of a system to save energy.
While some approaches dynamically exploit existing load imbalances to reduce the energy consumption of parallel applications~\cite{Rountree2009}, others try to find an energy-optimal operating point of the hardware during the whole program run~\cite{Mijakovic2016}.

The Horizon 2020 project READEX developed a semi-automatic tool suite for optimising the energy efficiency of scientific applications.
In its basic form, the tool suite estimates optimal configurations for different program regions before the application is run in production.
READEX tunes various hardware and software parameters, like traditional core DVFS or the uncore frequency, which controls the speed of non-core components of the processor~\cite{Hackenberg2015}.
The exact configuration for each of the program regions is stored in a configuration file.
During production runs, this file is used to adjust the processor frequencies accordingly.

While this approach enables users to improve the energy efficiency for different programs, it still has shortcomings.
Depending on the processed data, the optimal operating point of a program region can change.
Furthermore, if users move to a different HPC system, their software needs to be re-evaluated.
Therefore, we extended the READEX tool suite with a self-tuning approach, which we present in this paper.

Modern processors provide power measurement interfaces like RAPL, which is supported by contemporary Intel processors~\cite{Hackenberg2015}.
In READEX, we leverage the data that stems from such sources to measure the energy consumption of program regions.
To improve the energy efficiency of regions, we apply Q-Learning techniques to detect and adapt to changes in the optimal operating point.
By using this method, we can adjust to a continually changing environment in terms of changing workloads and HPC systems.

The paper is organised as follows: \secref{related} gives a short overview of other work using self-tuning with a focus on energy efficiency.
In \secref{readex} the relevant details of the READEX tool suite are explained.
The employed algorithm is explained in \secref{approach}.
In \secref{result}, we demonstrate our approach using the Kripke benchmark.
\secref{summary} summarises our findings and provides an outlook.

\section{Related Work}
\label{sec:related}

While self-tuning for energy efficiency is a broadly studied topic, the approaches used by researchers vary.
For example, Dhiman et.\,al.~\cite{Dhiman2009} use online learning to choose one of two energy-saving strategies (experts) during runtime.
They specify two experts: one for Dynamic Power Management (DPM) and one for DVFS.
The DPM expert decides whether to shut down a specific device during an idle period.
The DVFS expert uses a model based on processor events to estimate the optimal frequency for the CPU.
Both experts are evaluate at every OS scheduler tick.
However, this approach does not necessarily fit HPC workloads, which are usually long running applications without extended idle times.

Another example is given by Shen~et.\,al.~\cite{Shen2012}, who use Q-Learning to find a trade-off between temperature, performance, and energy.
They define a state as a combination of instructions per second, stall cycles, the current processor frequency, and the current processor temperature.
This state is captured and evaluated every 10\,ms.
Based on the observations, they calculate a reward, which is used to update the Q-value.
Based on the information, an action is taken, i.e., the frequency is changed.
Shen~et.\,al. use models to evaluate the performance and energy impact of their decisions.
However, these models can result in wrong predictions and have to be carefully chosen.
By using well-defined program regions, we can measure the energy and performance impacts, which is more precise than a model evaluated every 10\,ms.

\section{The READEX Toolsuite - Basic Infrastructure}
\label{sec:readex}

READEX is based on two tools: Score-P~\cite{Knuepfer2012} and the Periscope Tuning Framework (PTF)~\cite{Mijakovic2016}.
Score-P is a performance monitoring framework, which uses automatic instrumentation of function calls, OpenMP regions, MPI calls to create performance profiles and traces.
However, with the extensions described in~\cite{Schoene2017}, the infrastructure can also be used for tuning.

\begin{figure}[b]
    \fbox{\includegraphics[width=.97\columnwidth]{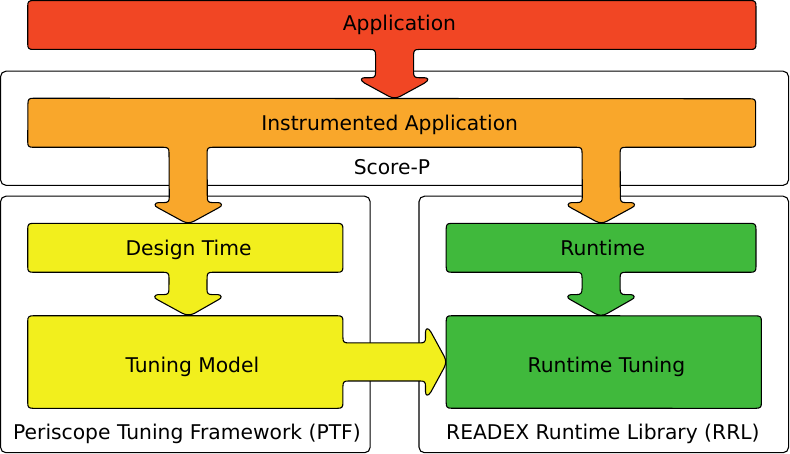}}
    \caption{
    The architecture of READEX.
    The application is instrumented using Score-P.
    Initially, a tuning model with the most energy efficient configurations is generated by PTF.
    For production runs, the tuning model is used by the RRL to apply configuration changes.
    }
\vspace{-1.5em}    
\label{fig:readex}
\end{figure}

The traditional READEX methodology consists of two steps:
(1) a \textit{design time analysis}, where the program and the energy efficiency of single regions is analysed, and
(2) the \textit{runtime tuning} where the gathered information is applied to tune single regions within the application.
Before the analyse step, the application has to be instrumented.
To limit the instrumentation overhead for short running functions, Score-P supports function filtering, which can be done during compile time or runtime.
READEX uses the compile time filtering method and selects only regions, with an average runtime of at least 100\,ms.
Afterwards, the program is analysed by PTF, which stores the optimal configuration for Runtime Situations (RTSs)~\cite{Schuchart2017} in a configuration file, the \textit{tuning model}.
The RTS approach extends the concept of an ``optimal configuration for a program region'' and puts regions into context by considering additional information when a region is called, e.g., the call-stack.
However, during runtime tuning, the instrumentation will remain intact and the tuning model is used by the READEX Runtime Library (RRL) to change the hardware and software environment according to the demands of the observed RTSs.
\figref{readex} shows an overview of the architecture.

Using this approach, the READEX tool suite achieves energy savings up to 15\%~\cite{Riha2018}.
Nevertheless, users need to generate a new tuning model as soon as they switch to a new HPC system.
Moreover, if users provide non-representative data set during design time, the applied configuration might be sub-optimal.
To mitigate these effects, we extend the RRL with a runtime detection mechanism for RTSs and a Q-Learning inspired algorithm to find the optimal configuration point, which we present in this paper.

\section{Self-Tuning Approach}
\label{sec:approach}
Q-Learning is an iterative algorithm that traverses over neighbouring states in an environment and evaluates them.
Usually, a Q-Learning problem reaches a final state and is then started again.
This, allows the algorithm to explore all different states and find an optimal path to the final state.

As an example, consider cliff-walking~\cite[p. 108 f.]{Sutton2010} where an agent tries to reach a goal without falling from a cliff.
Initially, the agent has no information about the area.
It randomly chooses steps and reaches a final state: either its goal or falling from the cliff.
Afterwards, the algorithm restarts and executes another overall iteration.
Now, it uses the previously collected information to avoid the cliff.
Over time, the agent accumulates enough information to reach its goal.

In our approach, there is no final state.
Hence, the algorithm terminates as soon as the program finished execution.
If the application is restarted, it is not necessary but possible to reset the algorithm.
Here, a few cases can be differentiated, from which we implemented three: (Re-)Evaluate a program whenever it is executed, discarding all previously collected information, continue the started overall iteration of the last execution, and finally restart the overall iteration from the initial state, and reuse the information.
The last option is closest to Q-Learning.

According to our measurements, one overall iteration already leads to good results if there are enough repetitions of the optimised region.
Therefore, we call our algorithm ``Q-Learning inspired'' as it still uses well-established decision algorithms from Q-Learning.
We base our algorithm on the one proposed by Sutton~\cite{Sutton2010} and add a few modifications to improve the convergence, as discussed in the following.

\subsection{Changes to the RRL}
In a self-tuning scenario, users skip the step of creating a tuning model.
Therefore, the RRL has to be able to create RTSs for unknown functions instead of reading the available RTSs from the tuning model.
To do so, we introduce the notion of a call tree into the RRL.
Unlike a call stack, which only contains a stack of instrumented functions until the function under investigation, the call tree includes all instrumented functions and user parameters that the program encountered so far.

A node in such a tree is either a function or a user parameter, which can be specified using an API.
For each new user parameter and newly encountered function, the RRL adds a new child node, starting with the \texttt{main} function as a root node.
Each function that is encountered, will also be profiled for runtime.
The tree information is then used to create an RTS by reading the path from the current node to the root node.
To filter short running functions, the RRL checks whether the current runtime of the current function is higher than 100\,ms.
If this is the case, it will be processed further:
If the function represents a leaf-node, it will be considered for optimisation.
If it is an internal-node, the RRL will consider the current node as an RTS if the combined runtime of all child nodes, which run shorter than 100\,ms is larger than the combined runtime of all child nodes, which run longer than 100\,ms.

\subsection{The Learning Algorithm}
In the following we use a notion, which is similar to the one Sutton uses in~\cite[p. 107]{Sutton2010}.
Here, $S_t$ describes the state at time $t$, $A_t$ the action taken at time $t$, $Q$ the state-action map, also called Q-Value function and $R_t$ the reward at time $t$.
Observations are noted after an action has been taken with $t+1$.
$\alpha$ is used to describe the learning rate, while a discount factor $\gamma$ is introduced as well.
Using this notation, the state-action map is calculated as follows\cite[p. 107]{Sutton2010}:
\begin{align}
 Q(S_t,A_t) \leftarrow ~ & Q(S_t,A_t) + \\\nonumber
 & \alpha \left[ R_{t+1} + \gamma \max_{a\in \cal{A}}(Q(S_{t+1},a)) - Q(S_t,A_t) \right].
\end{align}

In our case, this can be interpreted as follows: Once the energy measurement is finished, the reward ($R_{t+1}$) is calculated.
The saved state and action is then used to retrieve the Q-Value which lead to the decision ($Q(S_t, A_t)$).
We consider the current core and uncore frequency as a state and the change to a new frequency as an action.
Moreover, the Q-Value of the next action that will be chosen is retrieved by evaluating $\max_a(Q(S_{t+1},a))$.
These values are used to update the Q-Value, which leads to taking action $A_t$.

We limit the possible exploration actions to the surrounding states of the current one, i.e., the next lower or higher core and uncore frequency step.
This leads to a continuous exploration of the available states.
The size of the resulting state matrix is $\{\mbox{\textit{core frequencies}}\}~x~\{\mbox{\textit{uncore frequencies}}\}$.
Each state has an associated action matrix of size $3x3$, which allows to increase, decrease, or persist both, core and uncore frequency.
The combination of state and action is called state-action map.

The action matrix is initialised with $0$, except for the current configuration, where it is set to $-0.1$.
This encourages the algorithm to explore new states, instead of remaining at the current configuration.
To speed up the convergence, we reuse previously gathered information for surrounding states to calculate the Q-Value for the current state.

As reward function, we use the normalised energy:
\begin{align}
 R = \frac{(E_t - E_{t+1})}{\frac{1}{2} (E_t + E_{t+1})}
\end{align}
Where $E_t$ is the energy measured for the previous state ($S_t$), and $E_{t+1}$ is the energy measured for the current state $S_{t+1}$.
The initial value $E_0$ is measured during the first time a region is encountered.

For following iterations, the algorithm will search for new states, as long as it can decrease the energy consumption.
If a decision leads to increased energy consumption, i.e., $E_{t+1}>E_{t}$ and $R<0$, the algorithm will return to the previous state, which leads to a positive reward.
If the energy consumption increases, i.e. $E_{t+1}<E_{t}$ and $R>0$, the algorithm will explore other directions, as going back would lead to a negative reward.
This leads to a local minimum in a reasonable time.

To find the global minimum, we introduce a $\varepsilon$-greedy policy.
With a probability of $\varepsilon$, the decision made by the algorithm is neglected, but a randomly chosen action is taken.

\begin{figure}[tb]
    \fbox{\includegraphics[width=.97\columnwidth]{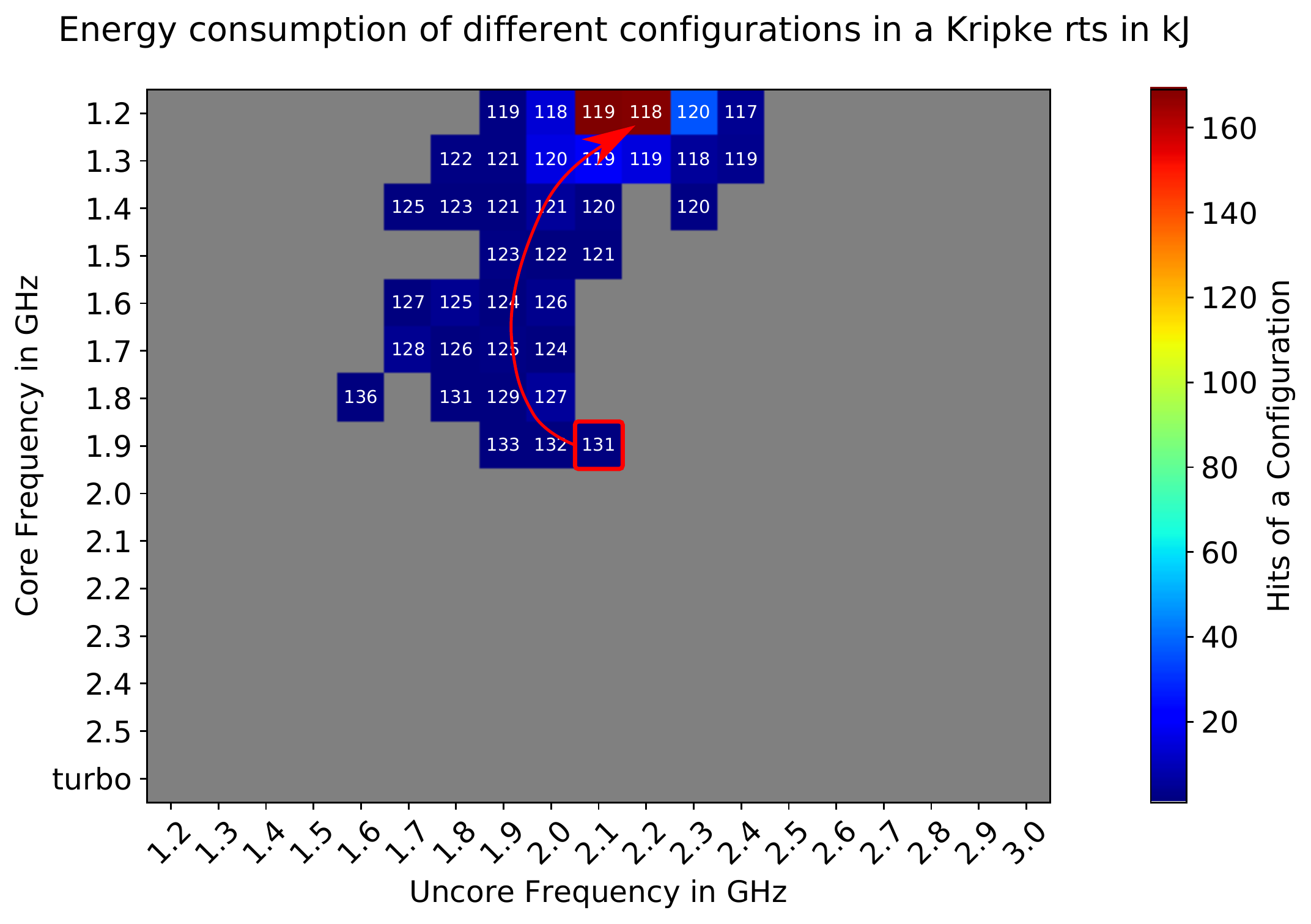}}
    \caption{
    The picture shows an RTS from Kripke, which is optimised using our algorithm.
    The red arrow indicates the overall optimisation direction.
    A minimum is reached in less than 50 steps.    
   While the colour indicates the iterations spend with a particular configuration, numbers indicate the last measured energy consumption at this specific configuration.
    }
\label{fig:kripke}
\end{figure}

\figref{kripke} illustrates the optimisation of an RTS using the described algorithm.
The algorithm starts at 1.9\,GHz core and 2.1\,GHz uncore frequency, and iterates until it finds a minimum at 1.2\,GHz core and 2.1/2.2\,GHz uncore frequency.

Coordinated decisions between different compute nodes would need communication between the various nodes.
As we can not guarantee that each process of an application runs the same functions in the same order, explicit communication might delay the execution or might even result in a dead-lock.
Therefore, we use local copies of the state-action map and the previously described call-tree and take decisions independently for each process.
The algorithm still converges to an optimal configuration in the long term.

\section{Results}
\label{sec:result}

To evaluate our algorithm, we choose the Kripke benchmark, as it provides the potential for dynamic energy savings~\cite{Vysocky2017}.
As Kripke supports OpenMP and MPI, we enable and instrument both.
We utilise a full compute node with OpenMP and use MPI only for the inter-node communication.
Using compile-time filtering, we reduce the instrumentation to only functions with more than 100\,ms runtime on one node.
We use $\alpha = 0.1$, $\gamma = 0.5$, and $\varepsilon = 0.25$, as they proved to be a good starting point.
However, it might be worth to investigate the effect of these hyperparameters further.

We use up to 24 nodes from the Taurus HPC systems installed at Technische Universit{\"a}t Dresden (TUD), Germany.
Each node is equipped with two Intel Xeon E5-2680 v3 processors, with 12 cores each, 64GB of RAM, and the HDEEM energy measurement system~\cite{hdeem}, which we utilise to evaluate of the full application run.
RAPL is used for decision making during the learning algorithm.
To account for the power consumption of the mainboard, network, and other components, we add an experimentally identified offset of 70\,W to the runtime measurements taken from RAPL.
We repeat each measurement four times.
The standard derivation is less than 1\% of the median energy.

\begin{figure}[tb]
	\fbox{\includegraphics[width=.97\columnwidth]{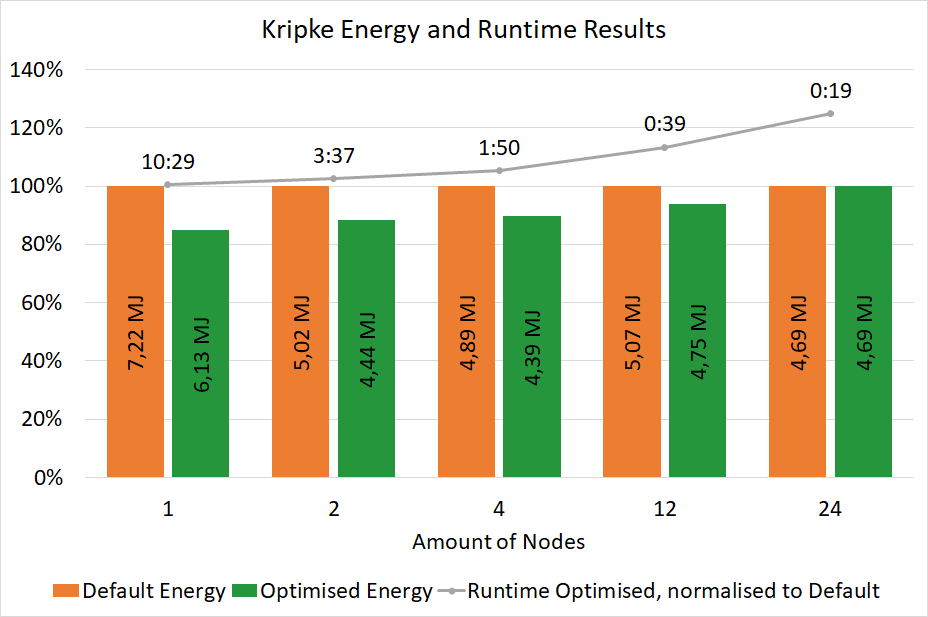}}
	\caption{
	The plot shows the results achieved using our algorithm.
	The values noted at the grey line indicate the runtime using our algorithm in hours and minutes, while the line itself shows the runtime normalised to the default runtime.
	The superlinear speedup from 1 node, with a runtime of 10:29, and 2 nodes, with a runtime of 3:37 might be related to the activation of MPI and a resulting better load balancing.
}
\label{fig:results}
\end{figure}

\figref{results} presents the results for the Kripke benchmark.
Using only one node, we can achieve up to 15\% energy savings, with only a 1\% increase in runtime.
However, the potential energy savings decrease as the node count increases.
It is evident that the increased runtime prevents any energy savings.
A more detailed analysis of the application, using Score-P's tracing capabilities, reveals that OpenMP and MPI functions with less than 100 ms are still present.
These cannot be filtered easily without possibly corrupting the information needed to create profiles and traces for parallel applications.
Each of these regions causes a call to Score-P, which might explain a portion of the overhead.
Moreover, the self-tuning happens independently for each process.
It is likely that different exploration decisions introduce load imbalances or slow down the convergence speed, which leads to a further increase in runtime.
Please note that the result for one node is close to what we reported using the READEX tool suite~\cite{Riha2018}, while no offline tuning with a representative workload is required.

\section{Conclusion and future work}
\label{sec:summary}

During the Horizon 2020 project READEX, we developed a semi-automatic energy efficiency tuning for HPC applications.
In this paper, we use the experience and concepts introduced by READEX and extended them with a self-tuning approach inspired by Q-Learning.
We were able to achieve improvements in energy efficiency for a small node count.
Further work is needed to filter program parts that are not relevant to optimisation, especially OpenMP and MPI calls.
Moreover, synchronised self-tuning should be evaluated.
Here RDMA could be used to synchronise the state-action map between all processes, as RDMA does not require explicit communication.
Finally, additional measurements are needed to evaluate if reusing the stored state action map leads to faster convergence and better energy savings for different workloads, as Kripke has a high dependency on the workload \cite{Vysocky2017}.

\section*{Acknowledgments}
\small
This work is supported in part by the German Research Foundation (DFG) within the CRC 912 - HAEC and by the European Union’s Horizon 2020 program in the READEX project (grant agreement number 671657).
Moreover, we like to thank Tobias Fischer for his support.

\bibliographystyle{IEEEtran}

\bibliography{paper}

\end{document}